\documentclass[aps,prb,showpacs,superscriptaddress,twocolumn]{revtex4}

\usepackage{epsfig}
\usepackage{amsmath,amssymb,amsthm}
\usepackage{graphicx}
\usepackage{psfrag}
\usepackage{bm}

\def\s2{\sigma^2}

\begin{document}

\title{Revivals, classical periodicity, and zitterbewegung of electron currents in monolayer graphene}

\author{E. Romera}
\affiliation{Departamento de F\'isica At\'omica, Molecular y Nuclear,
Universidad de Granada, Fuentenueva s/n, 18071 Granada, Spain
}
\affiliation{Instituto Carlos I de F{\'\i}sica Te\'orica y
Computacional, Universidad de Granada, Fuentenueva s/n, 18071 Granada,
Spain}

\author{F. de los Santos}
\affiliation{Instituto Carlos I de F{\'\i}sica Te\'orica y
Computacional, Universidad de Granada, Fuentenueva s/n, 18071 Granada,
Spain}
\affiliation{Departamento de Electromagnetismo y F{\'\i}sica de la
Materia,
Universidad de Granada, Fuentenueva s/n, 18071 Granada,
Spain}

\date{\today}

\begin{abstract}
Revivals of electric current in graphene in the presence of an
external magnetic field are described. It is shown that when the electrons are
prepared in the form of wave packets assuming a Gaussian population of
only positive (or negative) energy Landau levels, the presence of the magnetic field
induce revivals of the electron currents, besides the classical cyclotron motion.
When the population comprises both positive and negative energy Landau levels,
revivals of the electric current manifest simultaneously with zitterbewegung and the
classical cyclotron motion. We relate the temporal scales of these three effects and
discuss to what extent these results hold for real graphene samples.

\end{abstract}
\maketitle

\section{Introduction}
Graphene, a single-atom layer of carbon atoms arranged in a honeycomb lattice, has attracted enormous attention recently.
It exhibits remarkable mechanic and thermal properties \cite{mech}, but the most studied aspect are its startling electronic properties.
Graphene can carry high current densities, its resistivity is less than that of silver \cite{resistivity} (the lowest resistivy material at room temperature),
and when exposed to a magnetic field at sufficiently low temperatures it shows an anomalous quantum Hall effect \cite{qhe}.
Many of these properties are  attributed to graphene's peculiar band structure,
quasi-free electrons in single-layer graphene propagating as massless Dirac particles \cite{massless}, which  opens the possibility of
probing relativistic quantum mechanics effects in condensed matter systems. For example, perfect tunneling of electron wave packets through
potential barriers, the Klein paradox, unobservable in particle physics, has been experimentally confirmed in graphene very recently \cite{kp}.
Additionally, interference between positive and negative frequency components of wave packets superposition of both positive and negative energy eigenstates
induces a highly oscillatory motion,
known as zitterbewegung (ZB), that appears as damped, rapid oscillations of free Dirac particles around their otherwise rectilinear average trajectories.
ZB has been the subject of intense analytical investigations, although direct observations
remain elusive due to the enormous characteristic frequency and the small amplitude of this motion.
However, ZB may influence quantities other than the average position of free Dirac particles \cite{examplesZB},
e.g. the nontrivial behavior of the conductivity in graphene \cite{katsnelson}, and
it has been argued that graphene in a magnetic field is a promising system for the experimental observation of ZB
\cite{rusin1}.

In this paper we describe how, depending on the electron wave packet's initial conditions, electric current regeneration occurs in graphene
under an external magnetic field, due to revivals and fractional revivals of the wave function.
In particular, it is shown that when the electrons are prepared in the form of wave packets assuming a Gaussian population of
only positive (or negative) energy Landau levels, the presence of the magnetic field
induce revivals of the electron currents, besides the classical cyclotron motion.
When the population comprises both positive and negative energy Landau levels,
revivals of the electric current manifest simultaneously with ZB and the classical cyclotron motion.
Electric currents due to ZB of electrons in graphene have already been described \cite{rusin1,rusin2}, but the occurrence of revivals,
a phenomenon purely quantum mechanical in origin, has not been reported yet despite that these oscillations are
slow enough and large enough to be detected.
Next, we briefly review the phenomenon of wave packet revivals and then study how they show up in graphene.

The time evolution of wave packets, relativistic and nonrelativistic, can be quite complex
due to quantum interference. However, several types of periodicity may emerge depending on the character of
the energy eigenvalue spectrum. If the initial wave packet is a superposition of eigenstates $\varphi_n(x)$ sharply
peaked around some large central $n_0$, different time scales can then be identified from the coefficients of the Taylor expansion
of the energy spectrum $E_n$ around the energy $E_{n_0}$ \cite{parker,averbukh},
\begin{equation}
E_n \approx E_{n_0}+E'_{n_0} (n-n_0)+\frac{E''_{n_0}}{2} (n-n_0)^2 +\cdots.
\label{expansion}
\end{equation}
For instance, propagating wave packets initially evolve quasiclassically and oscillate
with period $T_{\rm Cl}=2\pi\hbar/|E'_{n_0}|$, which is in accordance with the correspondence principle.
Wave packets then spread and collapse, and the classical oscillations eventually damp out.
At later times, multiples of the revival time $T_{\rm R}=4\pi \hbar/|E''_{n_0}|$, collapsed wave packets (almost) regain their
initial waveform and oscillate again with period $T_{\rm Cl}$.
Moreover, at times that are rational fractions of $T_{\rm R}$, wave packets split into a collection of scaled and
reshifted copies called fractional revivals. Longer time scales can be defined beyond $T_{\rm R}$, v.g. the so-called superrevival
time, at which a new cycle of full and fractional revivals commences again.
Lastly, the term $E_{n_0}$ in the expansion Eq. (\ref{expansion}) above just generates an unobservable overall phase (see, however, below).
Revival phenomena have attracted considerable interest over the past decades. In particular, revivals
and fractional revivals have been investigated theoretically in nonlinear quantum systems, atoms and molecules \cite{mondragon},
and observed experimentally in, among others, Rydberg wave packets in atoms and molecules, molecular vibrational states, and Bose-Einstein
condensates \cite{rev_exp}. Interestingly, methods for isotope separation \cite{isotope}, wave packet control \cite{wp_control},
as well as for number factorization \cite{primenumbers} have been put forward that are based on revival phenomena.

\section{Revivals of the electric current in graphene}
To study wave packet revivals in graphene, we consider the Hamiltonian for electrons in graphene under a magnetic
field perpendicular to the plane and of intensity $B$.
Following standard methods \cite{rusin1}, within the tight binding approximation and near the vicinity of the ${\bf K}_1$ point, one of the two
inequivalent corners of the Brillouin zone, the Hamiltonian reads (we use units such that $\hbar=1$)
\begin{equation}
H_1=v_F \begin{pmatrix} 0& {\hat \pi_x} - i{\hat \pi_y}\\ {\hat \pi_x} +
  i{\hat \pi_y} & 0 \end{pmatrix},
\label{hamiltoniano}
\end{equation}
where $\hat{\pi}={\bf k} -e{\bf A}$ is the quasiparticle momentum, $\bf A$ is the vector potential and $e$ the electron charge.
The electronic dispersion relation is given by $E({\bf k}) \approx \pm v_F |\bf k|$, where ${\bf k}=(k_x,k_y)$ is the momentum measured relatively to
the ${\bf K}_1$ point and $v_F\simeq 10^6$ m/s is the Fermi velocity (see \cite{wallace,castro-guinea} for details).
Upon introducing the magnetic radius $L=\sqrt{1/eB}$ and using the Landau
gauge ${\bf A}=(-By,0,0)$, it is not difficult to show that the energy eigenfunctions are given by
\begin{equation}
\varphi_{k_x,n,s}(x,y)=\frac{e^{ik_x x}}{\sqrt{4\pi}} \begin{pmatrix}  -sf_{n-1}(\xi) \\ f_{n}(\xi)  \end{pmatrix},
\end{equation}
with $n=0,1,\cdots, s=\pm 1$ for the conduction and valence bands, respectively, and $k_x \in \mathbb{R}$. Here,
$f_n(\xi)=e^{-\frac{1}{2}\xi^2} H_n(\xi)/\sqrt{L}C_n$, $\xi=y/L-k_x L$, $C_n=\sqrt{2^nn! \sqrt{\pi}}$,
and $H_n(\xi)$ are the Hermite polynomials. The energy spectrum is, in turn, $E_{n,s}=sE_n=s \Omega \sqrt{n}$,
with $\Omega=\sqrt{2} v_F/L$.

We shall construct the initial wave packets as the linear combination
\begin{equation}
\Psi(x,y)=\int dk_x \sum_{n,s} c_{n,s}(k_x) \varphi_{k_x,n,s}(x,y).
\label{wp1}
\end{equation}
Two different situations will be considered. First, wave packets with $s=1$,
centered around given $k_{0x}$ and $n_0$ with coefficients $c_{n,s=-1}(k_x)=0$ and $c_{n,s=1}(k_x)=c_n(k_x)$
Gaussianly distributed as
\begin{equation}
c_{n}(k_x)=
\sqrt{\frac{d_k}{\pi\sqrt{\sigma}}}e^{-d_k^2(k_x-k_{ox})^2/2}
e^{-\frac{(n-n_0)^2}{2\sigma}}.
\label{gaussian_cn}
\end{equation}
The classical period and the revival time yield straightforwardly
$T_{\rm Cl}=4\pi \sqrt{n_0}/\Omega$ and $T_{\rm R}=16 \pi n_0^{3/2}/\Omega $, respectively.
In this case, ZB is not possible because  the wave packets contain positive energy states only.
Secondly, we shall consider a superposition state of two wave packets with opposite $s$ and centered around the same $k_{0x}$ and $n_0$ as before,
with coefficients $c_{n,s=-1}(k_x)=c_{n,s=1}(k_x)=c_n(k_x)$ Gaussianly distributed as given by (\ref{gaussian_cn}).
In this second case, simultaneously with the classical periodicity and revivals, ZB shows up with a period $T_{\rm ZB}=  \pi /E_{n_0}= \pi/\Omega n_0^{1/2}$
that is obtained from the first term in the expansion Eq. (\ref{expansion}), which  no longer acts as an unobservable overall phase.

\begin{figure}
\includegraphics[width=8cm]{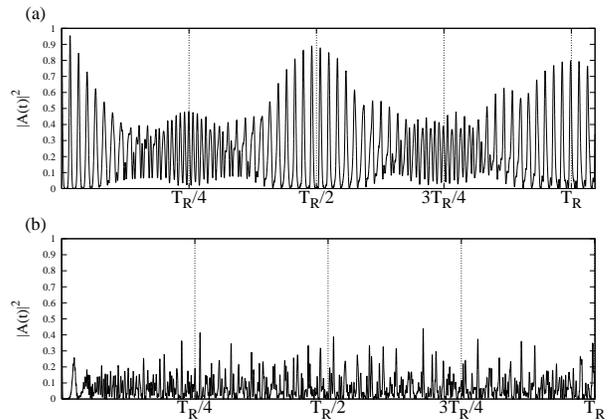}
\caption{Time dependence of $|A(t)|^2$ for initial Gaussian wave packets with $B=10$ T and (top panel) $n=15, \sigma=3$, $T_{\rm R}\simeq17$ ps
and (bottom panel) $n=11,\sigma=40$, $T_{\rm R}\simeq 11$ ps.}
\label{autocorrelation}
\end{figure}

To account for the regeneration of the time-evolving wave packets we shall use the autocorrelation function \cite{rob}
\begin{eqnarray}
A(t) &= &\int dx dy \Psi^{*}(x,y,0)\Psi(x,y,t).
\end{eqnarray}
An alternative approach based on information entropies has been recently proposed \cite{nosotros}.
After defining
\begin{equation}
U_{m,n}=\int_{-\infty}^\infty c^*_m(k_x) c_n(k_x) dk_x = \frac{1}{\pi\sigma} e^{-\frac{(n-n_0)^2}{2\sigma}}e^{-\frac{(m-n_0)^2}{2\sigma}}
\end{equation}
one readily finds
\begin{equation}
 A(t)=\sum_{n,s} U_{n,n} e^{-iE_{n,s}t}.
\end{equation}
The occurrence of revivals corresponds to the
return of $|A(t)|^2$ to its initial value of unity, and the fractional revivals
to the appearance of relative maxima in $|A(t)|^2$.  This is shown in figure \ref{autocorrelation} for $B=10$T and two different initial wave packets. In the top panel, where $n_0=15$, $\sigma=3$, and $s=1$, revivals can be clearly seen at $T_{\rm R}\simeq 17$ ps and  $T_{\rm R}/2$, and fractional revivals at $T_{\rm R}/4$ and  $3T_{\rm R}/4$.
In the bottom panel, however, upon changing to $n_0=11$ and $\sigma=40$, no wave packet regeneration takes place because of the smaller $n_0$ and large $\sigma$.

\begin{figure}
\includegraphics[width=8cm]{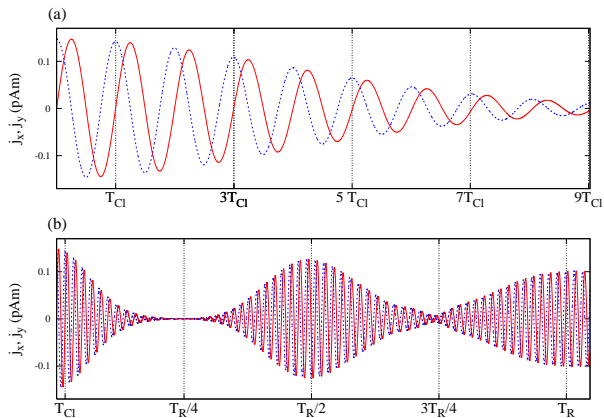}
\caption{
Time dependence of electric currents $j_x$ (blue, dotted line) and $j_y$ (red,
solid line) in graphene for $B = 10$ T, $n_0 = 15$, $\sigma = 3$, and $s=1$.
(a) First classical periods of motion with $T_{\rm Cl}\simeq 279$ fs. (b) Long-time dependence with $T_{\rm R}\simeq 17$ ps.
The classical periods and the main fractional revivals are indicated by vertical
dotted lines.
}
\end{figure}

\begin{figure}[b]
\includegraphics[width=8cm]{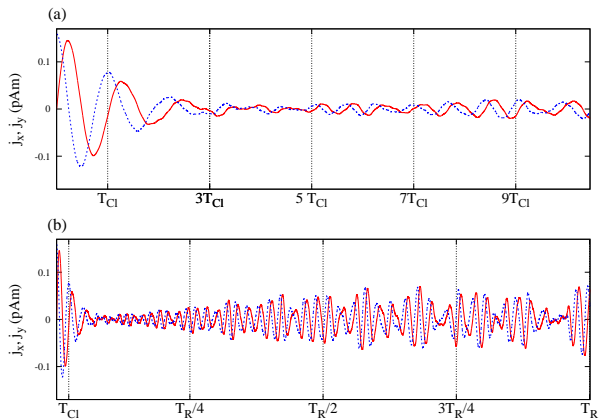}
\caption{Same as in Fig. 2 but with $n_0=11$ and $\sigma=40$.
$T_{\rm Cl}\simeq 239$ fs and $T_{\rm R}\simeq 11$ ps.
}
\end{figure}

To investigate the behavior of electron currents, we first compute the $x$ and $y$
components of the current. The electron velocity operators are given by
$v_j=i[H,r_j]/\hbar=v_F \sigma_j,(j=x,y)$,
where $\sigma_x$ and $\sigma_y$ are the Pauli matrices. When only positive ($s=1$) or negative ($s=-1$)
Landau energy levels are populated, upon expanding in the base $\varphi_{k_x,n,x}$ one finds for the expected temporal evolution of the currents
\begin{eqnarray}
j_x \equiv -e v_x(t) &=& e s v_F \sum_{n=1}^\infty U_{n-1,n} \cos \left[(E_n-E_{n-1})t\right],\nonumber \\
j_y \equiv -e v_y(t) &=& e v_F \sum_{n=1}^\infty U_{n-1,n} \sin \left[(E_n-E_{n-1})t\right].
\end{eqnarray}
The main features of the time evolution of $j_x$ and $j_y$ are extracted  by using the expansion  Eq. (\ref{expansion}). One
finds $E_n-E_{n-1} \approx E_{n_0}^{\prime}+ E_{n_0}^{\prime\prime} (n-n_0)$ from which the scales
$T_{\rm Cl}$ and $T_{\rm R}$ arise, in agreement with our previous discussion.
It is shown in figure 2a how at early times, for $n_0=15$ and $\sigma=3$, $j_x$ and $j_y$ follow fairly well
the classical cyclotron motion with a period $T_{\rm Cl}\simeq 279$ fs.
In a few periods the wave packet enters the collapse phase and the quasiclassical oscillatory behavior of the currents vanishes,
only to emerge later at half the revival time (see Fig. 2b). A revival of the electric currents can be observed at a time about $T_{\rm R} \simeq 17$ ps.
By contrast, for $n_0=11$ and $\sigma=40$,  the quasiclassical behavior
vanishes right away (Fig. 3a), and the wave packet does not regenerate at long times (Fig. 3b) due to insufficient localization.
Notice that for revivals to be neatly observed, there is no need to reach very large values of $n_0$ as long as $\sigma$ is sufficiently
decreased.

\begin{figure}
\includegraphics[width=8cm]{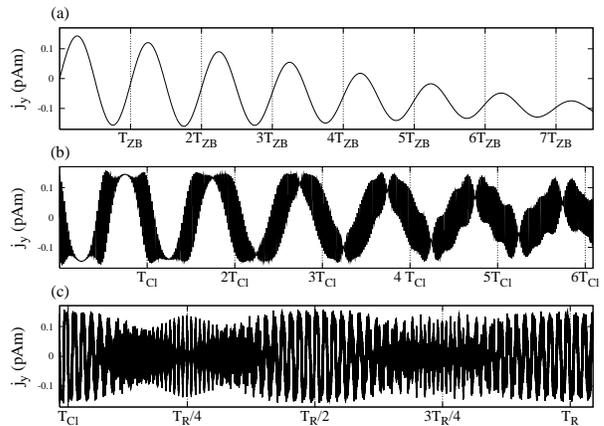}
\caption{
Time dependence of electric current $j_y$ in graphene for $n_0=15$,
$\sigma=3$, and $s=\pm 1$. Rest of parameters as in Fig. 2. The vertical
dotted lines stand for (a) ZB of electrons with $T_{\rm ZB} \simeq 4.7$ fs, (b) classical motion, and (c) revival behavior.
}
\end{figure}

We now consider an initial superposition state of two wave packets with
$n_0=15$, $\sigma=3$, $s=1$, and $s=-1$. It is straightforward to obtain
\begin{eqnarray}
j_x(t) &=& 0 \nonumber \\
j_y(t) &=& 2e v_F \sum_{n=1}^\infty U_{n-1,n} \{ \sin \left[(E_n+E_{n-1})t\right] \nonumber \\
&& \hspace{2.6cm}  +\sin\left[(E_n-E_{n-1})t\right]\}.
\end{eqnarray}
In this case, three different types of oscillatory motion arise. At very short times (in
the femtosecond scale), the electronic current is affected by ZB, as shown in
Fig. 4a. Using again the expansion Eq. (\ref{expansion}) leads to
$E_n+E_{n+1} \approx 2 E_{n_0}$, from which $T_{\rm ZB}=2\pi /2E_{n_0}=\pi/\Omega n_0^{1/2} \simeq4.7$ fs time-scale arises,
which coincides with the expression previously derived on more general grounds.
At medium times (Fig. 4b), ZB oscillations can be seen superimposed on the clearly visible, quasiclassical ones.
Similarly, full and fractional revivals can be identified in the
picosecond scale (Fig. 4c). Observe$c_{n,s=-1}(k_x)=0$ that, contrarily to what one would expect,
the fractional revivals are clearly seen, i.e. they are not blurred due to the presence
of two wave packets rather than a single one.
Turning to $n_0=11$ and $\sigma=40$, ZB persists but the 
quasiclassical periodicity fades away after the first three classical periods
(Fig. 5a), and the revivals are completely missed (Figs. 5b and 5c).

\begin{figure}
\includegraphics[width=8cm]{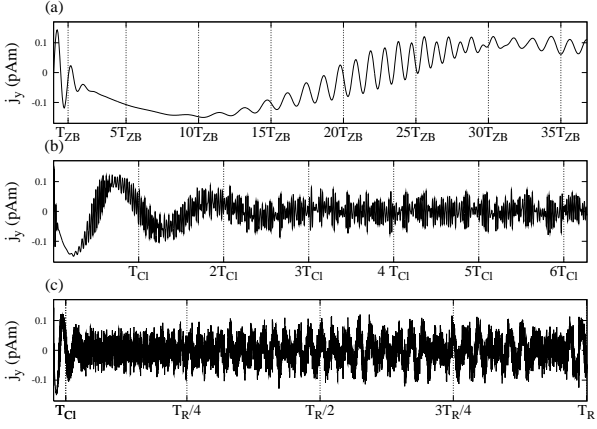}
\caption{Time dependence of electric current $j_y$ in graphene for $n_0=11$,
$\sigma=40$, and $s=\pm 1$. Rest of parameters as in Fig. 2. The vertical
dotted lines stand for (a) ZB of electrons with $T_{\rm ZB}\simeq 5.4$ fs, (b) classical motion, and (c) revival behavior.
}
\end{figure}

We now turn to examine to what extent the above predictions hold for real graphene samples.
It is well known that, in addition to the ${\bf K}_1$ point, there is a second inequivalent point
in the Brillouin zone, ${\bf K}_2$, where the valence and conduction bands meet.
The Hamiltonian for electrons at the ${\bf K}_2$ point is given by \cite{rusinnuevo,bena}
\begin{equation}
H_2=v_F \begin{pmatrix} 0&  -{\hat\pi_x} - i{\hat \pi_y}\\ -{\hat \pi_x} +
  i{\hat \pi_y} & 0 \end{pmatrix},
\label{hamiltoniano2}
\end{equation}
and hence $H_2=-H_1^{T}$. The eigenvalues are the same than those of $H_1$ but the
eigenfunctions are different,
\begin{equation}
\xi_{k_x,n,s}(x,y)=\frac{e^{ik_x x}}{\sqrt{4\pi}} \begin{pmatrix}  f_{n}(\xi) \\ sf_{n-1}(\xi)  \end{pmatrix},
\end{equation}
and the quantum velocity is given by $v_x=-v_F \sigma_x$ and $v_y=v_F
\sigma_y$. Electrons around the ${\bf K}_1$ and ${\bf K}_2$ points 
are excited to levels associated with $\varphi$ and $\xi$, respectively. Therefore,
the total electric current is the sum of two contributions coming from two different wave packets:
$\Psi$ as given by Eq. (\ref{wp1}) and $\Phi$ obtained after replacing 
$\varphi_{k_x,n,s}$ by $\xi_{k_x,n,s}$ in Eq. (\ref{wp1}), the coefficients
$c_{n,s}(k_x)$ being common to both wave packets (see \cite{rusinnuevo}) because 
the energy spectra of both Hamiltonians are identical. The  contribution
to the electric current from these two wave packets is equal \cite{rusinnuevo}.

\begin{figure}
\includegraphics[width=8cm]{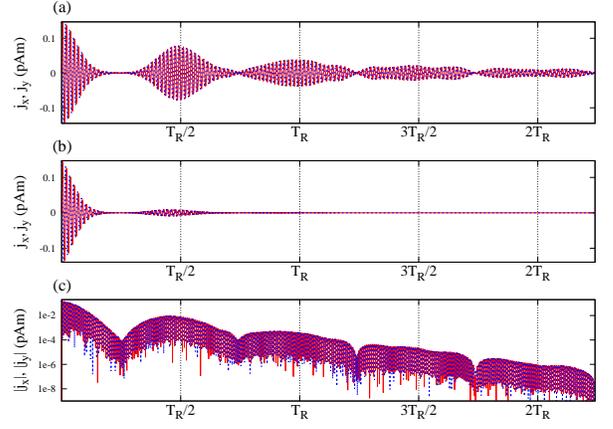}
\caption{Influence of the broadening of the Landau levels for $n_0=15$ and $B=10$ T.
Rest of parameters as in Fig. 2. For  $\Gamma=0.7 \ {\rm meV} < \Gamma_{\rm max}$ 
the revival behavior is still observable (panel (a)). For $\Gamma= \Gamma_{\rm max}$ 
most of the revivals are missed (panel (b)), but revival behavior can still be appreciated  in a logarithmic scale (panel (c)).
}
\label{broadening}
\end{figure}

Another feature of real graphene that might have an impact on revival phenomena
is the existence of a small gap, of the order of $10$ meV, caused by
impurities \cite{peres}, $e-e$ interactions, etc. \cite{Li,vozmediano}. However, the periods $T_{\rm Cl}$ and $T_{\rm R}$
will be the same than those calculated in the gapless model due to the fact
that they are derivatives of the energies, and only the $T_{\rm ZB}$ period will
change quantitatively with a new interband frequency $\hat{\Omega}=(\Omega^2+E^2_{gap})^{1/2}/\hbar$ \cite{rusinnuevo}.
A second consequence of the imperfections in graphene is the broadening of the Landau levels,
which may strongly influence the revivals of the electric current calculated using idealized, delta-like
Landau levels. Assuming finite widths for all energy levels, characterized by
broadening parameters $\Gamma_n$, the Landau levels $E_n$ are
replaced by complex energies $\tilde{E_n}=E_n+i \Gamma_n$
\cite{rusinnuevo}. The temporal evolution of the currents can then 
be calculated as
\begin{eqnarray}
j_x  &=& e s v_F \sum_{n=1}^\infty U_{n-1,n} \cos
\left[(E_n-E_{n-1})t\right] e^{-(\Gamma_n+\Gamma_{n-1})t},\nonumber \\
j_y  &=& e v_F \sum_{n=1}^\infty U_{n-1,n} \sin \left[(E_n-E_{n-1})t\right] e^{-(\Gamma_n+\Gamma_{n-1})t}.
\label{gammaonj}
\end{eqnarray}
We have estimated the maximum width $\Gamma_{\rm max}$ up to which revivals of the electric current are clearly observable 
for $n_0=15$ and $B=10$ T. Using the approximation $\Gamma_n\cong\Gamma$ for $n$ around
$n_0$, we obtain $\Gamma_{\rm max} \lesssim 3.7$ meV.
The influence of the broadening of the Landau level on the revivals of the electron current is illustrated in Fig. \ref{broadening}.
For $\Gamma=0.7 \ {\rm meV} < \Gamma_{\rm max}$ the revivals  are still observable (panel (a)), but for $\Gamma= \Gamma_{\rm max}$ 
most of them damp out due to the exponential factor in Eq. (\ref{gammaonj})
(panel (b)). The early time surviving revivals with $\Gamma= \Gamma_{\rm max}$ are visible 
when the current is plotted in a logarithmic scale (panel (c)). Therefore, the occurrance of revivals of the
electron current depends critically on the broadening of the Landau levels.

\section{Discussion and conclusions}
In conclusion, we have studied the dynamics of electron currents
in graphene subject to a magnetic field. Several types of periodicity
must be distinguished if the wave packets representing the electrons are sufficiently
localized around some large enough central quantum number $n_0$.
In this case, currents initially evolve quasiclassically and
oscillate with a period $T_{\rm Cl}$, but at later times the wave packet
eventually spreads, leading to the  collapse of the classical oscillations.
At times that are multiple of $T_{\rm R}$, or rational fractions of $T_{\rm R}$,
the wave packet (almost) regains its initial form, and the electron current its initial amplitude.
For this to occur, the presence of a quantizing magnetic field is necessary,
for if $B=0$ the spectrum is continuous which rules out the possibility of revivals \cite{rob}.
Associated with the revival of the wave
packet the quasiclassical oscillatory motion of the currents resumes.
Additionally, when both positive and negative Landau levels are populated, permanent ZB oscillations
are observed, in agreement with previous results \cite{rusin1}. Like $T_{\rm Cl}$ and $T_{\rm R}$, $T_{\rm ZB}=\pi /E_{n_0}$ 
can be simply obtained from the expansion Eq. (\ref{expansion}), 
thus providing a unifying description of three main time scales for the study of electric currents in graphene.
Interestingly, although all the several periods depend on $B$ as $1/\sqrt{B}$, their ratios $T_{\rm Cl}/T_{\rm ZB}=T_{\rm R}/T_{\rm Cl}=4n_0$
only depend on $n_0$. Moreover, 
since $T_{\rm ZB}<T_{\rm Cl}<T_{\rm R}$ current revivals are more accessible to experimental probing.
In particular, for the values used in this work $T_{\rm R}/T_{\rm ZB}=16n_0^2
\sim O(10^3)$. Interestingly, the analogy of Hamiltonian (\ref{hamiltoniano}) with the Jaynes-Cummings
model allows us to relate ZB and Rabi oscillations, as other
authors have pointed out \cite{schliemann,dora,bermudez}.
All these results carry over without change when a more refined modeling of graphene is used that accounts 
for the ${\bf K}_2$ point of the Brillouin zone and the energy gap caused by inevitable imperfections and 
interactions. However, the broadening of the Landau levels destroys most of
the revivals in the case of $B=10$ T, for broadening parameters 
$\Gamma \geq \Gamma_{\rm max}= 3.7$ meV. Since current experimental estimates of the line width are of the
order of 5 meV for $B=10$ T, the observation of revivals of the electric current in graphene 
would require either stronger magnetic fields, which lead to lower $T_{\rm R}$ and line width \cite{Li}, 
or the use of high quality graphene samples. We hope the present work will stimulate further experimental
studies in this direction.

\section{acknowledgments}
We would like to thank two anonymous referees for their helpful comments on this paper.
This work was supported by projects FQM-165/0207 and FIS2008-01143.


\begin{thebibliography}{99}
\bibitem{examplesZB}
Possible indirect observations of ZB have been proposed in ion-trap systems
\cite{bermudez}, photonic crystals \cite{photonic_crystals} and ultracold
atoms \cite{ultracold_atoms}.
\end{thebibliography}
\end{document}